\documentclass[%
 reprint,
superscriptaddress,
amsmath,amssymb,
aps,
prb,
]{revtex4-2}

\newcommand{\Edit}[1]{\textcolor{black}{#1}}

\usepackage{graphicx}
\usepackage{dcolumn}
\usepackage{bm}
\usepackage{color}
\usepackage[normalem]{ulem}
\usepackage{hyperref}

\begin{document}

\preprint{APS/123-QED}

\title{Field-direction sensitivity of Kondo hybridization in UTe$_2$}

\author{Thomas Halloran}
\email{thallor1@umd.edu}
\affiliation{NIST Center for Neutron Research, Gaithersburg, Maryland\ 20899, USA}
\affiliation{Department of Physics and Astronomy, University of Maryland}

\author{Gicela Saucedo Salas}
\affiliation{NIST Center for Neutron Research, Gaithersburg, Maryland\ 20899, USA}
\affiliation{Department of Physics and Astronomy, University of Maryland}
\author{Sylvia K. Lewin}
\affiliation{NIST Center for Neutron Research, Gaithersburg, Maryland\ 20899, USA}
\affiliation{Department of Physics and Astronomy, University of Maryland}
\author{J. A. Rodriguez-Rivera}
\affiliation{NIST Center for Neutron Research, Gaithersburg, Maryland\ 20899, USA}
\affiliation{Department of Materials Science, University of Maryland, College Park, MD 20742}
\author{Colin L. Sarkis}
\affiliation{Neutron Scattering Division, Oak Ridge National Laboratory, Oak Ridge, TN 37831, USA}
\author{Jakob Lass}
\affiliation{PSI Center for Neutron and Muon Sciences, 5232 Villigen PSI, Switzerland}

\author{Daniel G. Mazzone}
\affiliation{PSI Center for Neutron and Muon Sciences, 5232 Villigen PSI, Switzerland}
\author{Marc Janoschek}
\affiliation{PSI Center for Neutron and Muon Sciences, 5232 Villigen PSI, Switzerland}
\affiliation{Physik-Institut, Universität Zürich, Winterthurerstrasse 190, CH-8057 Zürich, Switzerland}
\author{Nicholas P. Butch}
\affiliation{NIST Center for Neutron Research, Gaithersburg, Maryland\ 20899, USA}
\affiliation{Department of Physics and Astronomy, University of Maryland}

\date{\today}

\begin{abstract}
\Edit{Neutron scattering experiments on the spin-triplet superconductor UTe$_2$ have established that the dominant low-energy magnetic response is along Brillouin zone boundaries, resembling the magnetic susceptibility of narrow-gap interband excitations}. We report a study of the sensitivity of these excitations to magnetic field along the crystallographic $\hat{a}$-axis.  Up to fields of $\mu_0 H$=13 T, the maximal inelastic neutron spectral weight increases in energy transfer, with a pronounced increase in $d\hbar\omega\Edit{_{peak}}/dH$ near $\mu_0 H$=7 T.  This behavior parallels the field and temperature dependent features of the electrical resistivity that are associated with Kondo hybridization.  Our measurements suggest that $\hat{a}$-axis fields near $\mu_0 H$=7~T induce a change in the hybridization between heavy $f$-electrons and the bare conduction band.

\end{abstract}

\maketitle


\section{\label{sec:intro}Introduction}





Uranium and Plutonium 5$f$ systems exhibit a rich variety of phenomena at low temperatures due to the interplay of interactions such as spin-orbit coupling, magnetism, and ill-defined valence occupancy. In many cases, the screening of localized paramagnetic moments by conduction electrons gives rise to a Kondo lattice and heavy-fermion superconductivity. Some examples of these effects can be found in UCoGe~\cite{Huy_2007,Stock_UCoGe_2011}, UPt$_3$~\cite{DeVisser_Menovsky_Franse_1987}, and URu$_2$Si$_2$~\cite{Broholm_1991,Butch_2015}. 

UTe$_2$ is a recently-discovered member of this family of materials, where it has been proposed that the itinerant spin-triplet superconductivity with $T_c\approx2.1~K$ is realized by magnetic interactions between spins~\cite{Ran_Butch_2019,Wei_Kapitulnik_2022}. The material has a number of unusual properties, and the precise nature of its superconductivity is an ongoing matter of research. Great progress has been made in recent years in the experimental study of the electronic band structure through photoemission measurements~\cite{miao_low_2020}, as well as in the characterization of the Fermi surface through high field quantum oscillation measurements~\cite{weinberger_quantum_2024, broyles_revealing_2023}. Progress has also been made from a theoretical perspective, with DFT+DMFT~\cite{chen_quasi-two-dimensional_2024,mekonen_optical_2022,miao_low_2020,kang_orbital_2022} works and emerging theories phenomenologically explaining the reentrant high field superconductivity observed experimentally~\cite{Hakuno_2024,Yu_Yu_Agterberg_Raghu_2023,Machida_2024}. 

To further these efforts, direct measurements of the magnetic interactions between spins are essential. Alongside tools like NMR~\cite{Nakamine_2021,ambika_possible_2022,Matsumura_2025} and magnetization~\cite{tokiwa_stabilization_2022,Lewin_Czajka_Science_2025} measurements, neutron scattering is a powerful method that directly probes the dynamical and static spin-correlation function. As no long-ranged magnetic order has been observed in the superconducting ground-state of UTe$_2$, all magnetic scattering is in the inelastic channel. This has been shown in all previous neutron studies, which report no elastic magnetic scattering and spin fluctuations that coincide with the Brillouin zone boundary~\cite{Knafo_2021,Butch_2022,Duan_2020,duan_resonance_2021}. These measurements are well described by magnetic moments that are preferentially oriented along the crystalline $\hat{a}-$axis with dynamical correlations at nonzero wavevectors. Our recent work has shown that these correlations are insensitive to $\hat{c}-$axis fields~\cite{Halloran_ute2_2025} up to magnitudes of $\mu_0H_c=13~T$, consistent with excitations originating from electronic interband excitons across the Kondo hybridization gap. Such excitons have been observed through neutron scattering in other materials like CePd$_3$~\cite{goremychkin_coherent_2018}.

To test this idea further, we performed inelastic neutron scattering measurements in a different configuration with the applied magnetic field along the crystalline $\hat{a}$-axis rather than the $\hat{c}$-axis. As the maximum field-scale accessible at scattering facilities is typically $|\mu_0H|\leq14~T$, the field-polarized and exotic high-field reentrant superconducting states are inaccessible in all field-directions~\cite{Lewin_Czajka_Science_2025}. However, the $\hat{a}-$axis low temperature susceptibility in UTe$_2$ is greater than the $\hat{c}-$axis susceptibility by a factor of approximately 5 times and the superconducting state is suppressed with an $\hat{a}$-axis field of $\mu_0H_a=9$ T (in $T_c>2~K$ samples) for $T$=0.3~K~\cite{Lee_Rosa_2025}, presenting an opportunity to explore potential field-induced effects in UTe$_2$. 

In contrast to the results for $\hat{c}$-axis fields, the scattering measurements reveal that the excitation spectrum is indeed modified with applied fields, which we take as evidence of band hybridization sensitivity to relatively small fields along the $\hat{a}$-axis. The evolution of the excitations with field is non-trivial, with a discontinuity being observed between fields of $5~T <\mu_0H<7~T$ and a change in $d(\hbar\omega_{peak})/dH$ for fields above and below this regime. \Edit{Thermopower, magnetization, and AC susceptibility} measurements have also reported anomalies in this field regime, which may be relevant to our results ~\cite{Niu_2020,tokiwa_stabilization_2022}. In general, the spectral weight shifts to higher energies \Edit{and the scattering decreases in intensity} with applied field, with the integrated intensity being suppressed to about half its original value at $\mu_0H_a=13~T$. The temperature-dependent electrical resistivity exhibits a similar anisotropic dependence on the applied magnetic field, reinforcing the connection between these phenomena and the effects of $f$-electron hybridization.

\section{Methods}

Two neutron scattering experiments were performed to determine the excitation spectra reported in this work. The experiments used different coaligned mosaics of single crystalline UTe$_2$. The mosaic used in the first experiment was previously used in Ref.~\cite{Halloran_ute2_2025}, where full details of synthesis and alignment may be found. 

The first neutron experiment was carried out on the cold multiplexing CAMEA spectrometer~\cite{Lass_2023} in the ($0kl)$ scattering plane at the Paul Scherrer Institute, Switzerland, with an applied magnetic field along the $\hat{a}$ direction. The total counting time was 30 hours at zero field, 20 hours at 11 T, 8 hours at 7 T, and 8 hours at 3 T. The experiment covered a range of energy transfers in $\hbar\omega\in\{0,6\}$ meV, with an instrumental resolution of $\delta_{FWHM}=0.11~$meV at the elastic line ($\hbar\omega=0$). The crystals were attached to a mount made of OHFC copper using CYTOP (AGC Chemicals Company) and coaligned using an X-ray laue diffractometer. All measurements were performed at a nominal temperature of $T=45(5)~$mK. All analysis was performed using the MJOLNIR software~\cite{Lass_2020}, and background subtraction was performed by selecting regions of $\bm{Q}-\hbar\omega$ space without magnetic scattering~\cite{lass_amber_2025}. The scattering has been normalized to absolute units using a vanadium standard of known mass.

A second mosaic used in the second neutron scattering experiment detailed below totals in mass of 1.15 g, with 104 coaligned single crystals of UTe$_2$. All crystals were grown using the molten salt flux method~\cite{Aoki_salt_2024,Sakai_salt_2022}, a distinction with prior experiments on samples made by chemical vapor transport. The critical temperatures of the batches of crystals used in this alignment were characterized by magnetometry and electrical resisitivty measurements, with $T_c\in\{1.95,2.05\}~K$, indicating high sample quality compared to samples used in older neutron experiments with typical $T_c\leq 1.7~K$. The crystals grown by the molten salt method offer two advantages: enhanced values of $T_c$, and crystal morphology. While crystals grown by chemical vapor transport tend to be rock-like, salt-flux crystals form rectangular platelets with a large facet normal the the $\hat{c}$-axis and the long edge being the crystalline $\hat{a}$-axis. This makes them more convenient to align, resulting in a tighter mosaic spread in the second experiment . 

The second neutron experiment was performed on the thermal time-of-flight SEQUOIA spectrometer at Oak Ridge National Laboratory~\cite{Granroth_2010}. The sample environment was a 14 T vertical magnet, with the sample mounted in the ($0kl$) scattering plane and the field applied along the $\hat{a}$ direction as in the CAMEA experiment. Measurements were performed with incident neutron energy of $E_i$=25 meV in the high resolution 240 Hz configuration, with a full-width resolution at the elastic line ($\hbar\omega=0~$meV) of $\delta_{FWHM}=0.8~$meV. All measurements were performed at the base temperature of $T$=1.9(1) K, with the exception of a $T$=60 K measurement used for background. \Edit{In previous studies, the magnetic excitations in UTe$_2$ have been shown to have little dependence on temperature between dilution temperature scales and $T=5~$K~\cite{butch_symmetry_2015,Knafo_2021}. Because of this, the scattering observed CAMEA and SEQUOIA is expected to be the same.} The inelastic scattering was measured for fields of 0 T, 5 T, 7 T, 11 T, and 13 T. The FWHM mosaic spread in the scattering plane was 3$^o$. The scattering was normalized to absolute units using integrated nuclear Bragg peak intensities, and details of the instrumental configurations for each scan along with counting times may be found in the Appendix. 

\section{Results}

\begin{figure}
    \centering
    \includegraphics[width=0.5\textwidth]{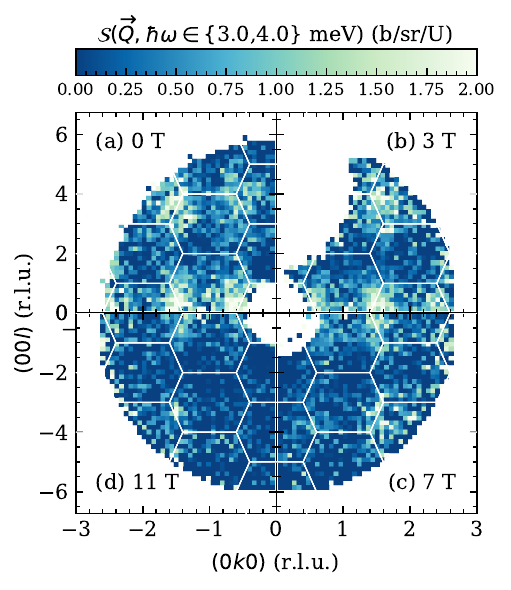}
    \caption{Inelastic neutron scattering from UTe$_2$ at $T=$45 mK in the $(0kl)$ scattering plane integrated in $\hbar\omega \in\{\Edit{3.0},4.0\}~$meV. Each quadrant shows the symmetrized scattering in $\hat{a}$-axis magnetic fields of magnitude 0 T (a), 3 T (b), 7 T (c), and 11 T (d). The scattering, which is strongest at Brillouin zone edges \Edit{which are denoted by white lines}, is substantially reduced by magnetic field along the $\hat{a}$-axis.}
    \label{fig:camea_constE}
\end{figure}

Field-dependent symmetrized constant energy inelastic neutron scattering intensity from the experiment on CAMEA is shown in Fig.~\ref{fig:camea_constE}. \Edit{As in previous studies, the scattering is centered at $\bm{Q}\neq 0$ wavevectors along the Brillouin zone boundary, with the maximal spectral weight being near energy transfers of $\hbar\omega\approx4$ meV at zero field}. All presented scattering intensity has been corrected by a non-magnetic background using the momentum averaged scattering from regions in reciprocal space where magnetic scattering from the sample is assumed to be absent, which is described in the Appendix. All presented measurements were corrected by the same background. The constant energy slice is integrated in energy transfer within $\hbar\omega\in\{\Edit{3},4\}~$meV which is the energy range of maximum scattering intensity at zero field, and the field evolution of the scattering intensity is clearly visible through the comparison of panels (a-d) (1 meV = 1.6x10$^{-22}$ J). In the available energy range, the scattering appears to almost vanish at the highest fields. This  differs significantly from the behavior in magnetic fields parallel to the $\hat{c}$-axis~\cite{Halloran_ute2_2025}, where the magnetic excitations in UTe$_2$ are independent of $\hat{c}$-axis fields up to 11~T.

\begin{figure}
    \centering
    \includegraphics[width=0.5\textwidth]{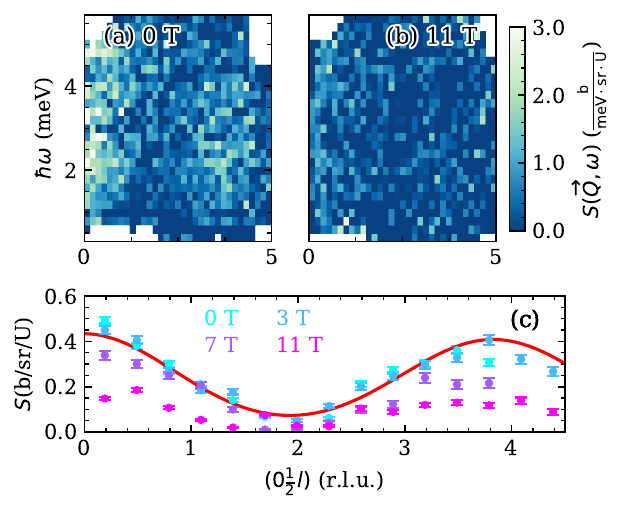}
    \caption{Symmetrized inelastic neutron scattering from UTe$_2$ along the $(0,\frac{1}{2}l)$ direction at zero field (a) and 11 T (b) from the CAMEA experiment. Here, the scattering has been integrated in $k\in\{0.3,0.7\}$ and $h\in\{-0.4,0.4\}$. In (c), the scattering intensity is shown as a function of $(0\frac{1}{2}l)$ in all four measured fields, showing that the excitations are peaked at $l=0$ and $l=4$. Scattering is integrated between $\hbar\omega\in\{1,4\}$ meV and $k\in\{1.25,1.75\}.$ The red line in (c) is a fit to two gaussian peaks centered at $l=0$ and $l\approx3.85$. \Edit{Both peaks are of the same amplitude, but modulated by the U$^{3+/4+}$ magnetic form factor. All error bars represent one standard deviation.}}
    \label{fig:camea_00L_slice}
\end{figure}

\begin{figure}
    \centering
    \includegraphics[width=1.0\columnwidth]{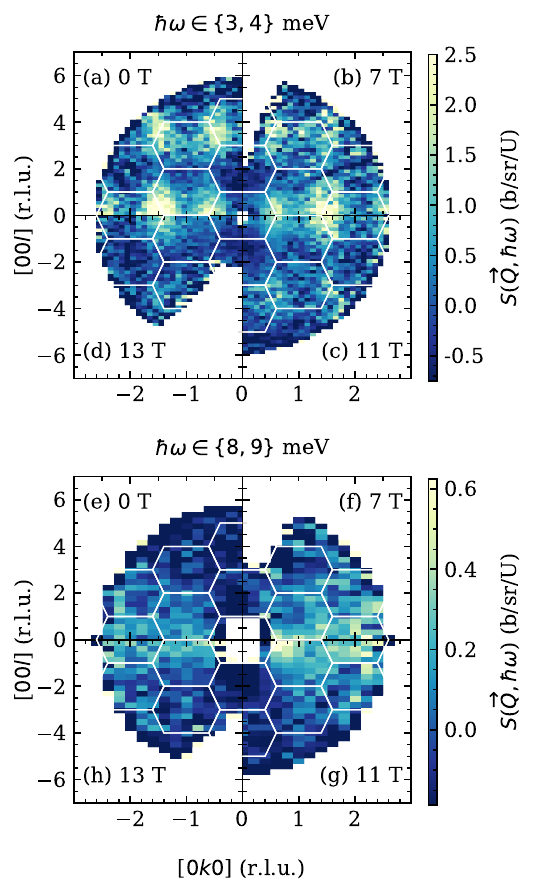}
    \caption{Inelastic neutron scattering from UTe$_2$ from second experiment on SEQUOIA at $T$=1.9(1) K, integrated in $\hbar\omega\in\{3,4\}$~meV in (a-d) and in $\hbar\omega\in\{8,9\}$~meV in (e-h). All slices have symmetrized and background subtracted as described in the appendix, and have been integrated in the out of plane scattering direction $h\in\{-0.5,0.5\}~$meV. Brillouin zone boundaries are shown as solid white lines, and the panels have fields applied along the crystalline $\hat{a}$-axis of (a,e) 0~T, (b,f) 7~T, (c,g) 11~T, and (d,h) 13 T}. 
    \label{fig:SEQ_constE}
\end{figure}

\begin{figure}
    \centering
    \includegraphics[width=1.0\columnwidth]{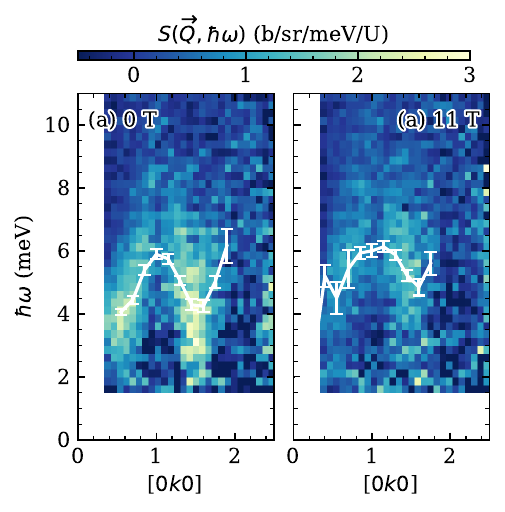}
    \caption{Dispersion of magnetic excitations from UTe$_2$ along the $(0k0)$ direction integrated in $h\in\{-0.4,0.4\}$ and $l\in\{-1,1\}$. Scattering has been background subtracted and symmetrized as described in the appendix. The white points are fits to better visualize the dispersion of the excitations at zero-field (a) and 11 T (b). Error bars represent one standard deviation uncertainty in the fitted dispersion energies. }
    \label{fig:SEQ_dispersion}
\end{figure}

\begin{figure}
    \centering
    \includegraphics[width=1.0\columnwidth]{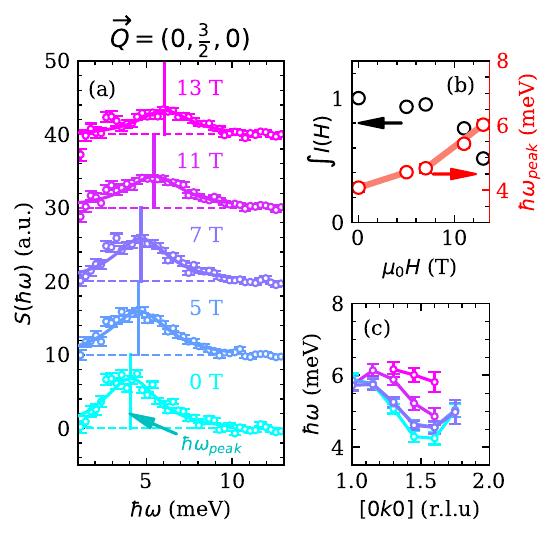}
    \caption{Summary of the evolution of the low-energy magnetic excitations in UTe$_2$ as function of magnetic field. (a) Energy-dependent spectra centered at the ($0,\frac{3}{2},0$), integrated between $h\in\{-0.4,0.4\}$, $k\in\{-1.7,-1.3\}$, and $l\in\{-0.5,0.5\}$. Intensities are shifted by 10 arbitrary units for each field, the solid lines represent fits to a Lorentzian form. The vertical solid lines denote peak positions as extracted from the fits. (b) Black points show the integrated intensity of the cuts in (a) normalized to the zero-field integrated intensity.  The red points show the shift in the peak in spectral weight as a function of field. \Edit{In this figure, error bars are smaller than the points but are included.} (c) Dispersion of magnetic excitations along the $(0k0)$ direction as a function of field, as extracted by fits to cuts along the energy direction. All error bars in (a) represent one standard deviation, and in (b,c) represent one standard deviation in fitted parameters.}
    \label{fig:SEQ_ecuts}
\end{figure}

The energy-dependent scattering is shown in Fig.~\ref{fig:camea_00L_slice} at zero field (a) and 11 T (b). As in previous studies~\cite{Duan_2020,Knafo_2021,Halloran_ute2_2025,Butch_2015}, there is no dispersion along the $l$-direction and the scattering intensity peaks at $\hbar\omega\approx$3 meV energy transfer. The scattering is periodic along the $(00l)$ direction, and the intensity fades with increasing magnetic field. The field-dependence of the intensity is highlighted in Fig.~\ref{fig:camea_00L_slice}(c), where the intensity in the $(0\frac{1}{2}l)$ direction has been integrated in energy within $\hbar\omega\in\{1,4\}$ meV. 

A second experiment focusing on higher energy transfers was conducted at the SEQUOIA instrument, with constant energy slices being shown in Fig. \ref{fig:SEQ_constE}. In Fig.~\ref{fig:SEQ_constE}(a-d), constant energy slices are shown in the \Edit{$(0kl)$} scattering plane which have been integrated between energy transfers of $\hbar\omega\in\{3,4\}$ meV, which may be directly compared to the first measurement. Fields of 0 T, 7 T, 11 T, and 13 T are compared, revealing the same trend as before with the main feature being a gradual reduction in scattering intensity with increasing field. However, as shown in Fig.~\ref{fig:SEQ_constE}(e-h), the trend at higher energy transfers ($\hbar\omega\in\{8,9\}$ meV) is different. Rather than \Edit{linearly} decreasing as a function of field like at low energy transfers, the scattering intensity increases upon applying a field of 7 T, remains enhanced at 11 T, then is suppressed at a high field of 13 T. This directly shows that the spectral weight does not vanish, as might be interpreted from Figure~\ref{fig:camea_constE}, but instead shifts to higher energy transfers with increasing magnetic field. 

To show this more clearly, we plot the dispersion of the magnetic excitations along the $(0k0)$ direction in Fig.~\ref{fig:SEQ_dispersion}. We compare two measurements, 0 T and 11 T, in Fig.~\ref{fig:SEQ_dispersion}(a) and Fig.~\ref{fig:SEQ_dispersion}(b) respectively. At zero field, the excitations are dispersive along the $(0k0)$ direction, as seen in previous works~\cite{Butch_2015,Halloran_ute2_2025}. The dispersions reported in these works are slightly different from that reported here due to a larger integration in the out of plane scattering along the $(h00)$ direction. This is shown more clearly through fits to the energy dependent intensity at specific points in $(0k0)$, which are shown by the solid white lines in Fig.~\ref{fig:SEQ_dispersion}. At 11 T in Fig.~\ref{fig:SEQ_dispersion}(b), the peak in spectral weight at the ($0\frac{3}{2}0)$ point shifts to higher energy transfer, going from $\hbar\omega=3.9(1)$ meV at 0 T to $\hbar\omega=5.2(1)$ meV at 11 T. In contrast, the spectral weight at the $(010)$ point does not shift, meaning that the bandwidth of the excitations narrows with applied field. 

The effect of field on the magnetic excitations is made more clear through one-dimensional cuts along energy transfer at the $(0\frac{3}{2}0)$ point, as shown in Fig~\ref{fig:SEQ_ecuts}(a). Each cut is fit to a Lorentzian form, from which the peak in spectral weight is extracted and represented by a solid vertical line. The integrated intensity of the excitations at each field is visualized in Fig.~\ref{fig:SEQ_ecuts}(b) by the black points, and is normalized to the maximal value at zero-field. Likewise, the peak in spectral weight is visualized using the red points in Fig.~\ref{fig:SEQ_ecuts}(b). 

An interesting result can be inferred from these \Edit{measurements}, \Edit{which is that the field evolution of the maximal spectral weight $d(\hbar\omega_{peak})/dH$ is not linear between zero-field and $\mu_0H_a=13~$T}. The rate of the evolution of the maximal spectral weight versus field $d(\hbar\omega_{peak})/dH$ approximately triples for the $\mu_0H\geq 7$ T regime when compared to the low-field regime as shown by the solid red lines in Fig.~\ref{fig:SEQ_ecuts}(b), going from $d(\hbar\omega_{peak})/dH$=0.073(2) meV/T to $d(\hbar\omega_{peak})/dH$=0.022(2) meV/T. Within the \Edit{uncertainty} of the least-squares fit parameters, the energy transfer of the maximal spectral weight in the 5 T $(\hbar\omega\Edit{_{peak}}=4.6(1)$ meV) and 7 T $(\hbar\omega\Edit{_{peak}}=4.7(1)$ meV) measurements are equal. This is supported by Fig.~\ref{fig:SEQ_ecuts}(c), which shows the magnetic dispersion of the excitations along the $(0k0)$ direction at every applied field. Indeed, the 5 T and 7 T scattering is identical not just at the $(0\frac{3}{2}0)$ point, but \Edit{within the full $(0k0)$ plane}. Apart from th\Edit{ese two fields,} which will be elaborated upon later in the text, it is clear that the \Edit{energy transfer associated with the maximal spectral weight} shifts to higher energies upon the application of increased $\hat{a}$-axis magnetic fields, and that the integrated intensity decreases to about half its zero-field value at 13 T. \Edit{Measurements at additional intermediate fields would be required to  verify the scattering is indeed field-independent between fields of $\mu_0H_a\in\{5~T,7~T\}$ This could readily be achieved using a triple-axis spectrometer fixed at the $\bm{Q}=(0,\frac{3}{2}{0})$ point.} Zero \Edit{magnetic} intensity is observed at energy transfers at $\hbar\omega>10$ meV, indicating that the measurement captures all of the magnetic intensity.
\begin{figure}
    \centering
    \includegraphics[width=0.5\textwidth]{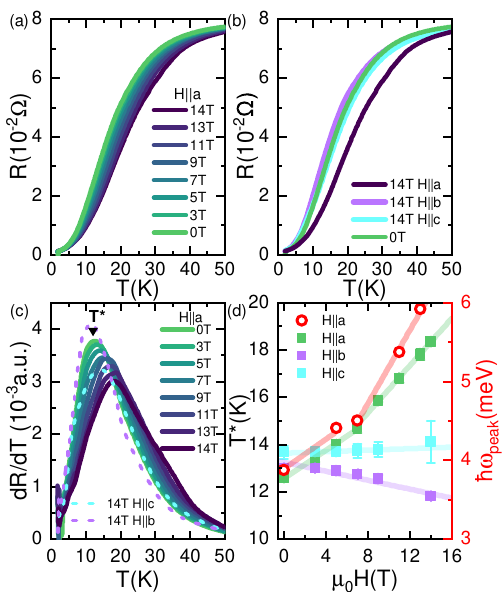}
    \caption{Temperature-dependent electrical transport measurements of UTe$_2$ with varying field. (a) Measurement along the $\hat{a}$-axis, showing that the inflection point shifts to higher temperature with increasing field. (b)  Comparison of the resistance at 14 T for the three principal axes with respect to 0 T, with the greatest shift being along the $\hat{a}$-axis. (c) Derivative of resistance with respect to temperature for all the fields measured along the $\hat{a}$-axis and  the 14 T measurement for fields along the $\hat{b}$ and $\hat{c}$-axes. (d) The quantity $T^*$, inferred from the maximal value of $dR/dT$ in (c) for is plotted for each axis. The extracted values of $T^*$ for each field show a positive slope for $H\parallel\hat{a}$. For $H\parallel\hat{b}$, the trend changes with a negative slope. For $H\parallel\hat{c}$, $T^*$ shows no significant change in field. Error bars represent one standard deviation of fitted parameters. The right axis of (d) plots the peak in spectral weight as extracted from neutron scattering in Fig.\ref{fig:SEQ_ecuts}. \Edit{The error bars in this figure are included, but are smaller than the point sizes.}}
    \label{fig:transport}
\end{figure}

Finally, we have performed electrical resistivity measurements from $T=1.8$ K to $T$=100 K with applied fields along the $\hat{a}$, $\hat{b}$ and $\hat{c}$ crystalline axes as shown in Fig.~\ref{fig:transport}. \Edit{In all cases, the applied current $\bm{J}$ was along the $\hat{a}$-axis.} The overall trend in the transport data as a function of $\hat{a}$-axis field is more clearly shown in Fig.~\ref{fig:transport}(d), which shows $dR/dT$ at each applied field. From this, it is clear that the inflection point in R(T) is shifting to high temperatures with increasing field strength. \Edit{Here we define the quantity $T^*$, which is the temperature of the maximal value of $dR/dT$.  This inflection point tracks the maximum value of R(T) but is a narrower feature and better defined. The maximum in R(T) is often associated with the coherence or Kondo temperature in other heavy fermion-systems such as CeCoIn$_5$~\cite{Bauer_2006} and URu$_2$Si$_2$~\cite{Butch_2015}.} At zero-field, this quantity is found to be $T^*=12.6(2)~K$, and increases with $\hat{a}$-axis field. If one examines the slope of the extracted values of $T^*$ with respect to the applied $H_a$, different values of $dT^*/dH_a$ can be extracted for fields above and below $H_a$=6 T. This, along with the field-evolution of the neutron scattering, directly evidences a nontrivial evolution of the electronic band structure that is worth further investigation in the intermediate field regime of $4~T\leq H_a \leq7~T$.

Another interesting aspect of the transport data is in comparison to the $\hat{b}$-axis and $\hat{c}-$axis field response. Along the $\hat{c}$-axis direction, the transport changes between zero-field and 14 T with a shift in $T^*$ of $\Delta T^*=1.6(9)$ K, which may be compared with the $\hat{a}-$axis shift of $\Delta T^*=5.9(1)~K$. This follows from our previous scattering results in the $\hat{c}-$axis field configuration which showed that the magnetic excitations are largely independent of $\hat{c}-$axis fields~\cite{Halloran_ute2_2025}. However, field applied along the $\hat{b}-$axis has the opposite effect, with a decrease in $T^*$ from $T^*$=12.6(2) K at 0 T to $T^*$=11.8(1) at 14 T, which is consistent with measurements reported in Ref.~\cite{Eo_2022}. Based on the correspondence between T* and $\hbar\omega_{peak}$, we expect the magnetic fields applied along the $\hat{b}$-axis will suppress $\hbar\omega_{peak}$.  

\section{Discussion}
To interpret the scattering, we first compare to the existing zero-field results in the literature. The spectra reported in this study is compatible with that found in previous works focused on the $(0kl)$ scattering plane in zero field~\cite{Knafo_2021,Duan_2020}. The scattering intensity at all fields is periodic in $(00l)$ as shown in Fig.~\ref{fig:camea_00L_slice}(c), which is well accounted for by the U$^{(3+,4+)}$ magnetic form factor and assuming that the nearest-neighbor U-dimers are in phase~\cite{Knafo_2021}. The scattering at low energies emerges from the corners of the Brillouin zone with a $\bm{k}=(0,0.6,0)$ wavevector, and has measurable dispersion along the $(0k0)$ direction as shown in Fig.~\ref{fig:SEQ_dispersion}, which was previously observed in Ref.~\cite{Butch_2022}. 

For fields along the $\hat{c}-$axis, our previous work~\cite{Halloran_ute2_2025} showed that the excitations are only weakly dependent upon magnetic fields up to field scales of $\mu_0 H_c=13~T$, in contrast to the $\hat{a}$-axis dependence observed in this study. The scattering appears to originate from interband excitations across the hybridization gap, as proposed in our previous work, suggesting that the band hybridization in UTe$_2$ is sensitive to $\hat{a}$-axis fields more so than $\hat{c}$-axis fields. It is also consistent with the bulk magnetic response, as the low temperature magnetic susceptibility is greater along the $\hat{a}$-axis than the $\hat{c}$-axis by a factor of five~\cite{Ran_Butch_2019}. 

For context, we turn to previous works on band structure instabilities in the related compound UCoGe. \Edit{Similar to UTe$_2$,} UCoGe is an Ising-type ferromagnetic superconductor that realizes rare reentrant superconductivity at high magnetic fields~\cite{hattori_superconductivity_2012,Huy_2007}. High field hall effect and thermopower measurements of UCoGe reveal that upon the application of a field along the magnetic easy-axis, and several Lifshitz transitions are observed which are induced by the polarization of magnetic spins~\cite{bastien_lifshitz_2016}. Here, it was proposed that \Edit{in the case of UCoGe} small Fermi surface pockets disappear with \Edit{each of the observed} Lifshitz transitions, thus changing the topology of the band structure, \Edit{which is ultimately reflected in the thermopower measurements of Ref.~\cite{bastien_lifshitz_2016}}.  

Similar measurements have been performed on UTe$_2$~\cite{Niu_2020}, \Edit{where the authors identify transitions that appear to be similar to that seen in UCoGe, with a particular transitino of interest at} $H_1=$5.6 T. \Edit{The nature of this transition is still an open question, but the authors of this work suggest this transition may be similar to those seen in UCoGe. To clarify if this is indeed a toplogical transition in the band structure, quantum oscillation measurements are required, as was done in the case of UCoGe~\cite{bastien_lifshitz_2016}.  Such studies are difficult in the case of UTe$_2$, where quantum oscillations have only been observed at fields well above the $\approx6$ T region of interest~\cite{weinberger_quantum_2024,broyles_revealing_2023}.}

Measurements of low-temperature magnetization with H$\parallel \hat{a}$ also reveal an apparent magnetic transition at approximately 6 T \cite{tokiwa_stabilization_2022}. This transition is argued to reinforce superconductivity and is correlated with a modification of the upper critical field. The authors of Ref. \cite{tokiwa_stabilization_2022} argue that there is an associated phase transition between two different superconducting order parameters. A similar interpretation is found in a recent NMR report revealing that the the spin degree of freedom within the spin-triplet pair can be manipulated by means of an applied external magnetic field \Edit{along the $\hat{b}$-axis}~\cite{Kinjo_2023}, thereby differentiating the high and low field superconducting phases. One might expect that this effect \Edit{would also be realized for fields} along the $\hat{a}$-axis.  Our neutron scattering and electrical resistivity measurements underscore that the 6 T transition modifies the electronic structure on a larger energy scale of the hybridization gap, namely 4 meV, but further studies are required to determine further details such as the change in the superconducting order parameter. 

The most direct follow up to this study would be a measurement of the evolution of the magnetic excitation in $\hat{b}-$axis fields. Based upon our transport data in Fig.~\ref{fig:transport}(d), the hybridization energy scale may actually decrease at accessible field scales in neutron scattering. Additionally, more detailed studies of the electronic Fermi surface in $\hat{a}$-axis fields, specifically in the intermediate field scale identified \Edit{in previous thermopower and electronic transport measurements as well as our current neutron study}, are necessary to understand the details of a \Edit{potential} band transition in this regime. One readily available measurement would be NMR within this transition regime, which has been done at slightly lower fields in Refs.~\cite{Matsumura_2025,Matsumura_2023}.

\Edit{We discuss our results in the context of superconductivity in UTe$_2$ and its relationship to applied magnetic field.  To review, even at modest strengths below 7 T, the magnetic field couples to the superconducting spin-triplet order parameter~\cite{Matsumura_2023,Nakamine_2021}, which could result in a superconducting phase transition in some cases. For even higher fields along the $\hat{b}$-axis, it is evident from bulk measurements that a first-order magnetic transition terminates superconductivity at 35 T~\cite{Lewin_2023,miyake2022magnetovolume,Schonemann2023}. This phenomenon presumably is associated with a change in the electronic structure, as opposed to the explicit field-stability of the superconducting phase, i.e., a spin or orbital limit. The resistivity data in Fig.~\ref{fig:transport} are consistent with a hybridization gap diminishing in field parallel to the $\hat{b}$-axis, and it is a reasonable extrapolation that the metamagnetic transition may be associated with a hybridization gap closure. Such a change in the electronic structure would have strong signatures in the inelastic neutron signal in a 35 T field along the $\hat{b}$-axis.  In contrast, the increased hybridization experienced for fields along the $\hat{c}$-axis does not obviously strengthen superconductivity.  This field-reinforced hybridization will not yield gap closure at high magnetic fields, which is consistent with the increasing values of the metamagnetic field upon rotation from the $\hat{b}$-axis to $\hat{c}$-axis~\cite{Lewin_Czajka_Science_2025}.}

\section{Conclusions}

In this work, we have presented a series of measurements of the excitations in UTe$_2$ in $\hat{a}$-axis magnetic fields up to field scales of $\mu_0H$=13 T. The excitations shift to higher energies with increasing field, with an exception between 5 T and 7 T where little field dependence is observed. We interpret this as evidence of the evolution of the hybridization gap in UTe$_2$, growing by $\approx2$ meV from that in the ground state at $\mu_0H_a=$13 T. This is supported by $T<$50 K electronic resistivity measurements along all principal crystal axes. \Edit{The magnetoresistance observed in these transport measurements is consistent with an increase in the magnitude of the Kondo hybridization gap in the influence applied magnetic fields along the $\hat{a}$-axis. This effect is not observed in fields along the $\hat{c}$-axis.} 

\Edit{This relates to the presented inelastic neutron scattering work, where the spectral weight shifts to higher energies in \textit{a}-axis fields but is robust against $\hat{c}$-axis fields, supporting an interpretation of the scattering as originating from interband excitations across the Kondo gap. We discuss the possibility of $\hat{a}$-axis magnetic field driven band instabilities.}

\section{Acknowledgments}
Identification of commercial equipment does not imply recommendation or endorsement by NIST. A portion of this research used resources at the Spallation Neutron Source, a DOE Office of Science User Facility operated by the Oak Ridge National Laboratory. The beam time was allocated to SEQUOIA on proposal number IPTS-34201. This work is based on experiments performed at the Swiss spallation neutron source SINQ (proposal number 20240248), Paul Scherrer Institut, Villigen, Switzerland. Work at UMD was supported by NIST and by the National Science Foundation under the Division of Materials Research Grant NSF-DMR 2105191.

\bibliography{refs}

\section{Appendix}

\section{\label{app:sample}Neutron Details}

Two different coaligned crystal assemblies were used in this work, the first being detailed in Ref.~\cite{Halloran_ute2_2025} and used in the CAMEA experiment and the second mosaic being a new one used in the SEQUOIA experiment. The new samples were synthesized by the molten salt method as detailed in Refs.~\cite{Aoki_salt_2024,Sakai_salt_2022} with all batches having superconducting critical temperatures $T_c \in \{1.95,2.05\}~K$. The second mosaic is shown in Fig., with the mount being made of aluminum. The samples were placed in a 14 T vertical magnet at the SEQUOIA beamline, with a maximal vertical divergence from the $(0kl)$ scattering plane of $\pm 10^o$ ($|h|< 0.4$ r.l.u for the primary experimental configuration of $E_i$=25 meV). 
\begin{figure*}
    \centering
    \includegraphics[width=1.0\textwidth]{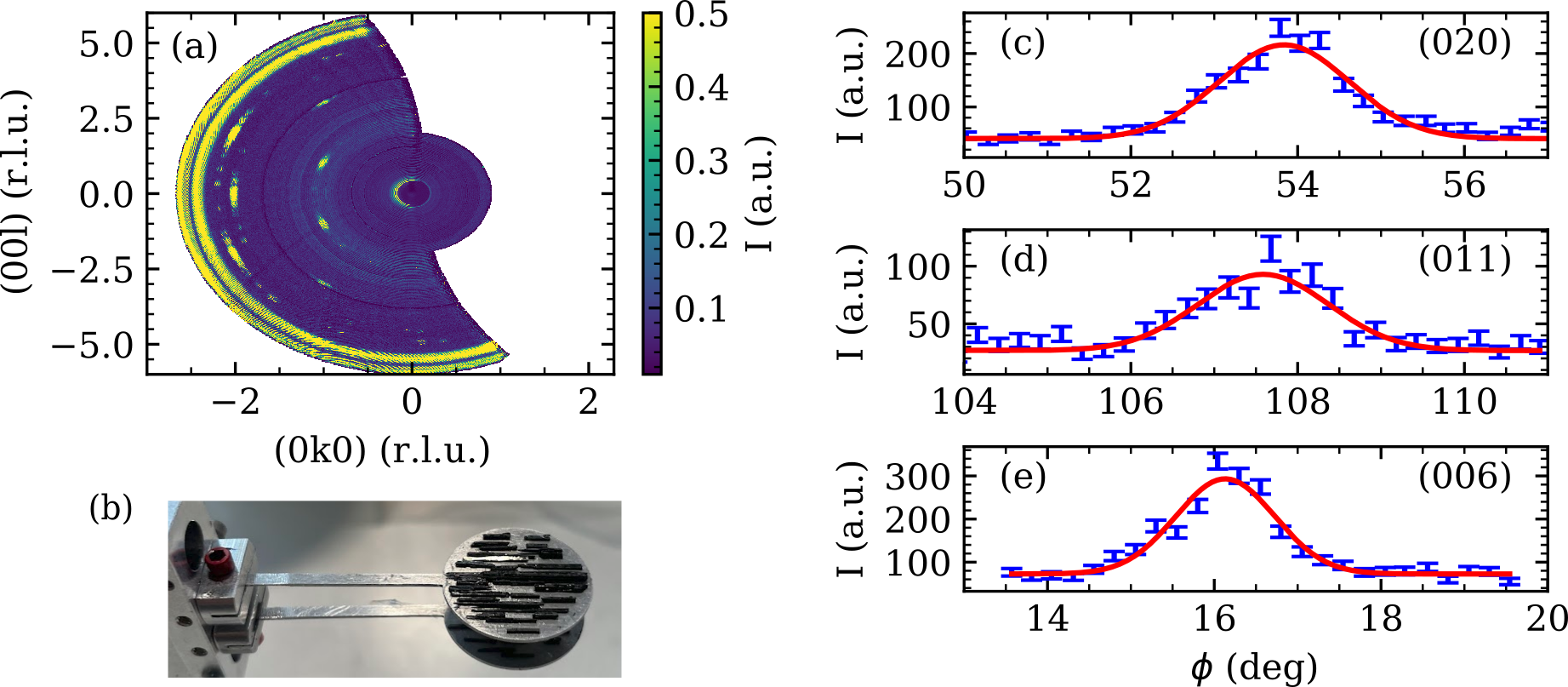}
    \caption{Characterization of mosaic of UTe$_2$ sample used in the SEQUOIA experiment. Elastic scattering integrated in $h\in\{-0.4,0.4\}$ and $\hbar\omega\in\{-1,1\}$ meV is shown in (a), showing nuclear Bragg peaks. A picture of the coaligned crystals and aluminum mount used in the experiment is shown in (b), and rocking curves are plotted in (c-e) for characteristic Bragg peaks are shown. The red lines are fits to a Gaussian form. From these curves the mosaic spread of $\delta_{FWHM}=3.5(2)^o$ was determined.}
    \label{fig:supp_mosaic_figure}
\end{figure*}

All scans were measured using E$_i$=25 meV neutrons and the high resolution 120 Hz chopper configuration. Measurements were performed $T=1.9(1)$ K apart from one at zero field and $T=60.0(1)$ K. The accumulated scattering time in each configuration was 76 C for $H=0$ T, 55 C for $H=5$ T, 58 C for $H=7$ T, 100 C for $H=11$ T, 38 C for $H=13$ T, and 76 C for the $H=0$ T and $T=60.0(1)$ K measurement. The sample was constructed using two circular plates, with an in-plane sample mosaic of $\delta_{FWHM}=3.5(2)^o$ as shown in Fig.\ref{fig:supp_mosaic_figure}.

Background subtraction of the CAMEA measurement was performed in the same manner as described in the appendix of Ref.~\cite{Halloran_ute2_2025}, using methods developed in Ref.~\cite{lass_amber_2025}. Specifically, the regions assumed to contain magnetic scattering are within $|l|\leq0.5$, $3.5\leq |l| \leq 4.5$ based on Refs.~\cite{Knafo_2021,Duan_2020}, using only the high field $\mu_0H=$11 T scattering. This region is used for all energies and values of $(0k0)$. 

Background subtraction of the SEQUOIA measurement was performed in a similar manner to the CAMEA data, but using $T=60.0(1)$ K scattering at zero field. In this case, regions of scattering were not isolated but instead specific sample rotation angles where magnetic scattering was minimal. The sum of these angles was then momentum-averaged and used as a background for all measurements using the SHIVER package~\cite{shiver}. 

The CAMEA scans were performed using E$_i$=3.8 meV, E$_i$=5.7 meV, E$_i$=7.7 meV, and E$_i$=9.36 meV for the zero field configuration, providing access to energy transfers of $\hbar\omega\in\{-0.5,8\}$ meV, with the total counting time being 52 hours. An identical measurement was performed with $\mu_0H$=11 T, with the total counting time being 32 hours. The scans at $\mu_0H=$3 T and $\mu_0$=7 T did not cover the full energy range, instead only focusing on the regime of peak intensity using E$_i$=5.7 meV and E$_i$=7.7 meV neutrons, with a total counting time of 15 hours and 12 hours for the $\mu_0H=$3 T and the $\mu_0H=$7 T configurations respectively.

\section{\label{app:fits}Dispersion Fitting}
In Fig.4 and Fig.5(c) of the main text we report the magnetic dispersion of the broad excitation spectra along the $(0k0)$ direction. This is done by performing a fit to a Lorentzian form of the background-subtracted data, integrating in bins of width 0.1 r.l.u. along the $[0k0]$ direction. Example fits are shown in Fig.~\ref{fig:supp_ecut_fits}, showing good agreement between the measurements and the fit. Points along $[0k0]$ where no sufficient fit could be found were excluded from these figures, where this is defined by the amplitude of the peak being three times its uncertainty.

\section{Field-dependent scattering at $(0,3/2,0)$}
\begin{figure}
    \centering
    \includegraphics[width=0.5\textwidth]{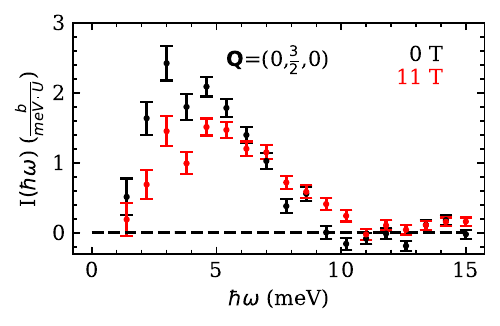}
    \caption{Scattering at the $(0,3/2,0)$ point demonstrating evolution of spectral weight from zero-field to $\mu_0H_a$=11 T.}
    \label{fig:SI_Ecuts}
\end{figure}

\Edit{It is unclear in Fig.~\ref{fig:SEQ_ecuts} the main text if the observed scattering intensity shifts to higher energies with field, or if intensity at low energies is suppressed and a high-energy part of the spectra remains static. To clarify this, we present scattering from the two highest-statistics datasets, $\mu_0 H$=0 T and $\mu_0 H=$11 T at the $(0,3/2,0)$ point as a function of energy transfer in Fig.~\ref{fig:SI_Ecuts} The scattering in this plot is integrated in $h\in\{-0.15,0.15\}$, $k\in\{1.3,1.7\}$, and $l\in\{-1.5,1.5\}$. From this plot it is clear that not only is scattering suppressed at low energies, but scattering intensity appreciably increases at higher transfers ($\hbar\omega\in\{7,10\}$ meV). It is unclear whether there are multiple components to this, i.e. a high and low energy peak splitting, which would require studies with more counting statistics.}

\begin{figure}
    \centering
    \includegraphics[width=0.5\textwidth]{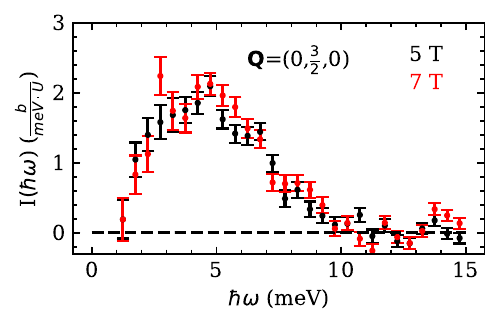}
    \caption{Scattering at the $(0,3/2,0)$ point demonstrating equivalence between $\mu_0H_a$=5 T and $\mu_0H_a=$7 T measurements. }
    \label{fig:SI_Ecuts_5T_7T}
\end{figure}

\begin{figure*}
    \centering
    \includegraphics[width=1.0\textwidth]{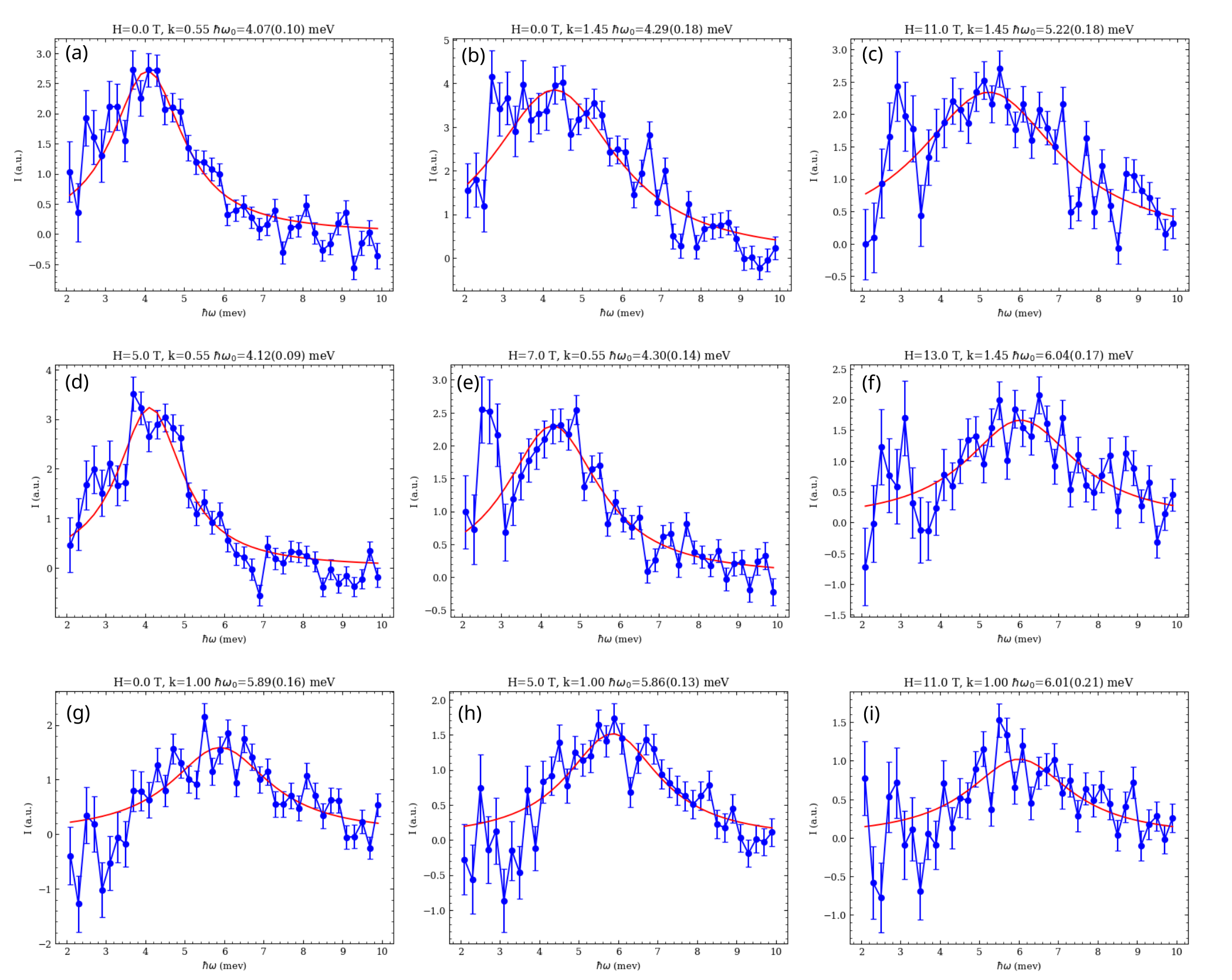}
    \caption{Representative fits of cuts along the energy axis used to determine dispersion in Figures~\ref{fig:SEQ_dispersion} and~\ref{fig:SEQ_ecuts}(c). Red line is a fit to a Lorentzian form with a constant background.}
    \label{fig:supp_ecut_fits}
\end{figure*}

Additionally, as there is discussion of the correspondence between the $\mu_0 H_a=5$ T and $\mu_0H_a$=7 T measurements in the main text, it is instructive to directly overlay these cuts along energy which is shown in Fig.~\ref{fig:SI_Ecuts_5T_7T}. In this figure, the integration regions are the same as Fig.~\ref{fig:SI_Ecuts}. Within the error from counting statistics, the two measurements appear to be the same. 

\section{\label{app:subsec}Electrical Resistivity Measurements}
\begin{figure*}
    \centering
    \includegraphics[width=1.0\textwidth]{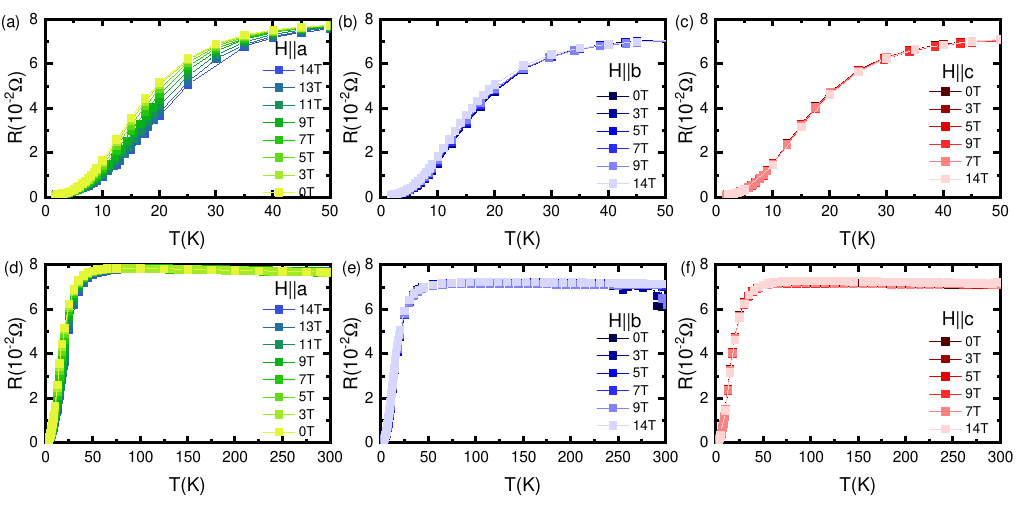}
    \caption{Electronic transport measurements f $T$=2 K to 50 K (a-c) and $T$=2K to 300 K (d-f) with fields along the primary axes as labeled. These measurements were used to produce Fig.~\ref{fig:transport}. The field-dependence of the transport is most clearly visible in (a), at temperatures below $T$=50 K along the $\hat{a}$-axis.}
    \label{fig:supp_transport_fig}
\end{figure*}

We performed field-dependent electrical resistivity measurements in a 14 T Quantum Design Physical Property Measurement System. Fig. A crystal grown by MSF of approximately 2 mm $\times$ 0.5 mm $\times$ 0.5 mm was used for resistance measurements with current applied along the $\hat{a}$-axis, with a standard four-point contact. Crystal orientation was confirmed using a Laue back-scattering system. Fig. \ref{fig:supp_transport_fig} presents all the data collected at all principal crystallographic axes. All error bars represent one standard deviation.

\begin{figure*}
    \centering
    \includegraphics[width=0.5\textwidth]{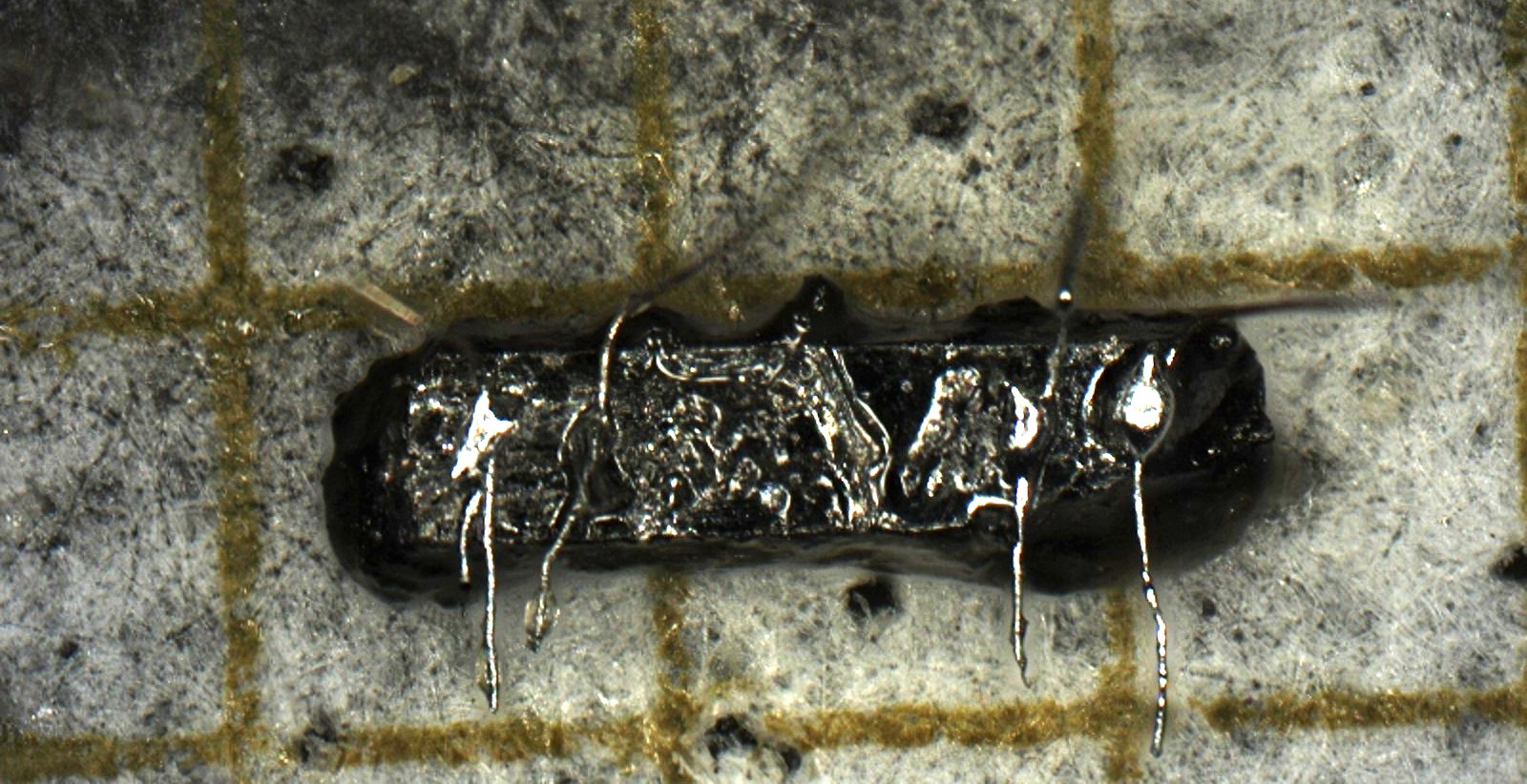}
    \caption{MSF crystal with leads along the $\hat{c}$-axis, as used in measurements shown in Fig.~\ref{fig:supp_transport_fig}(c,f).}
    \label{fig:supp_sample}
\end{figure*}

\end{document}